# Conductivity contrast and tunneling charge transport in the vortex-like ferroelectric domain patterns of multiferroic hexagonal YMnO$_3$

E. Ruff[1], S. Krohns[1], M. Lilienblum[2], D. Meier[2,3], M. Fiebig[2], P. Lunkenheimer[1,*], and A. Loidl[1]

[1]*Experimental Physics V, Center for Electronic Correlations and Magnetism, University of Augsburg, 86159 Augsburg, Germany*
[2]*Department of Materials, ETH Zurich, 8093 Zurich, Switzerland*
[3]*Department of Materials Science and Engineering, Norwegian University of Science and Technology, 7043 Trondheim, Norway*

We deduce the intrinsic conductivity properties of the ferroelectric domain walls around the topologically protected domain vortex cores in multiferroic YMnO$_3$. This is achieved by performing a careful equivalent-circuit analysis of dielectric spectra measured in single-crystalline samples with different vortex densities. The conductivity contrast between the bulk domains and the less conducting domain boundaries is revealed to reach up to a factor 500 at room temperature, depending on sample preparation. Tunneling of localized defect charge carriers is the dominant charge-transport process in the domain walls that are depleted of mobile charge carriers. This work demonstrates that via equivalent-circuit analysis, dielectric spectroscopy can provide valuable information on the intrinsic charge-transport properties of ferroelectric domain walls, which is of high relevance for the design of new domain-wall-based microelectronic devices.

The hexagonal manganites $R$MnO$_3$ ($R$ = Sc, Y, In, and Dy-Lu) form a unique group of multiferroics where a geometrically-driven mechanism triggers improper ferroelectricity [1]. Additional interest in this material class arose from the reported occurrence of vortex-like ferroelectric domain patterns [2,3,4]. Around the vortex cores, forming the centers of "cloverleaf" patterns of six domains, the polarization changes sign six times. These cores, evolving at a high-temperature structural transition [5], represent stable topological defects. Even strong electric fields only lead to a variation of ferroelectric domain sizes in these materials, but are unable to completely eradicate unfavorable domains and to generate a mono-domain state [2,6]. Moreover, there is a strict coupling of ferroelectric and antiferromagnetic domain walls (DWs), the latter forming at much lower temperatures around 100 K [7].

Recently it was shown that these complex domain properties also may be of relevance from an application point of view: Conductive atomic-force microscopy (c-AFM) on ErMnO$_3$ [4] and HoMnO$_3$ [8] revealed that the conductance of the ferroelectric DWs is either enhanced or suppressed compared to the domains, being determined by the polarization orientation of the adjacent domains. As the DWs can be easily tuned by external fields, this opens the possibility of domain-boundary engineering and applications in microelectronics using the nanoscale DWs instead of the domains themselves as active device elements [9,10,11,12,13]. The hexagonal manganites seem especially suited for this kind of functionality: Their DWs are robust and represent persistent interfaces as they are attached to the vortex cores, but within these constraints they can be moved by an external field thus enabling switching [4,12,13].

In general, insulating domain walls, also observed in various other systems as SrMnO$_3$ thin films [14] and (Ca,Sr)$_3$Ti$_2$O$_7$ [15], have shifted into the focus of interest, due to their possible applications, e.g., as rewritable nanocapacitors. The conductivity contrast between these DWs and the domains should be as high as possible. However, there is literally no data available that unambiguously proves the intrinsic nature of the reduced conductance and the mechanisms behind the residual domain-wall currents have not been tackled. For example, c-AFM, applied in Refs. [4] and [8] to hexagonal manganites, relies on the formation of Schottky-like barriers between the metallic tip and the semiconducting sample and thus essentially detects surface effects. Moreover, the current flow from the AFM tip to the bottom electrode leads to an ill-defined geometry. Thus, while the detected conductance variations of about one decade [4] or a factor of 4 [16] in ErMnO$_3$ and of 25% in HoMnO$_3$ [8] provide strong indications for different conductivities of domains and walls, no unequivocal information on the absolute value of the conductivity contrast is gathered. Recent photoemission electron-microscopy experiments on ErMnO$_3$ have demonstrated the intrinsic nature of the conductivity variation between domains and DWs [17] but this technique also does not provide absolute values. These are fundamental problems that go beyond hexagonal manganites and also apply to functional ferroelectric domain walls in other systems.

Thus, new and unambiguous insight into the local transport behavior is highly desirable. Interestingly, bulk dielectric spectroscopy is able to reveal information about the electrical properties of different regions in heterogeneous samples, even without any efforts to sense the behavior of specific sample regions by using microscopic contact geometries [18,19]. In the present work, we have employed dielectric spectroscopy to YMnO$_3$, probably the most studied hexagonal manganite, to determine the actual conductivity contrast between bulk (i.e., the domains) and domain boundaries with reduced conductance. There are several previous dielectric studies of this compound (e.g., [20,21,22,23]). Their results are inconsistent, which points to a strong influence of non-intrinsic effects and a considerable sample dependence of the dominant dielectric properties. We investigate two single crystals, measured as grown or subjected to different cooling rates after annealing above the structural phase transition ($T_c \approx 1260$ K [5]), leading to distinct vortex densities [24,25].

Single crystals of YMnO$_3$ were grown by the flux method (sample 1) and the floating-zone technique (sample 2) [5]. Their geometry is roughly plate-like with the $c$ axis vertical to the surface (sample 1: area 4.7 mm$^2$, thickness 0.3 mm; sample 2: 2.4 mm$^2$, 0.4 mm). Sample 1 was measured as grown and after annealing in N$_2$ gas at 1400 K (well above $T_c$ [5]), followed by cooling with 1 K/min down to 900 K to generate a defined vortex density (denoted as sample 1a) [24,25]. Sample 2 was annealed at 1420 K and cooled with 10 K/min. For the dielectric measurements, contacts of silver paint or sputtered platinum were applied to opposite faces of the crystals, ensuring an electrical field direction parallel to $c$. The dielectric constant $\varepsilon'$ and conductivity $\sigma'$ were determined using a frequency-response analyzer (Novocontrol Alpha-A Analyzer). Sample cooling and heating were achieved by a closed-cycle refrigerator and a home-made oven, with the sample in vacuum. For the determination of the vortex density, piezo-response force microscopy (PFM) was performed at room temperature [4].

Figure 1 shows $\varepsilon'(T)$ and $\sigma'(T)$ of sample 1 at various frequencies between 10 and 600 K. $\varepsilon'(T)$ [Fig. 1(a)] exhibits a peak reaching "colossal" [18] values up to $10^4$. Its left flanks shift with frequency, signifying a relaxational process, and the overall behavior resembles that of so-called relaxor ferroelectrics [26,27]. However, one should be aware that the well-known non-intrinsic Maxwell-Wagner (MW) relaxations arising in heterogeneous samples, in special cases can also mimic relaxor-ferroelectric behavior [18,28,29]. The low-temperature flanks of the $\varepsilon'(T)$ peaks seem to be composed of two consecutive steps indicating a large relaxor-like and an additional relaxation process with smaller amplitude, termed in the following relaxation 2 (at higher temperatures) and 1 (at lower temperatures), respectively. Relaxation steps are usually accompanied by peaks in the dielectric loss $\varepsilon''$, which is proportional to $\sigma'/\nu$. In $\sigma'(T)$ [Fig. 1(b)], only the smaller relaxation 1 is revealed by a peak. It only shows up as a shoulder (e.g., at about 200 K for the 1 Hz curve) due to the superposition by the non-zero conductivity of the semiconducting sample, which strongly increases with temperature.

The larger relaxor-like mode 2 most likely is of MW type, arising from a so-called surface-barrier layer capacitor (SBLC) at the sample surface [18,30]. The missing $\sigma'$ peak for this relaxation indicates similar conductance of the SBLC and the bulk [29]. Peaks in $\varepsilon'(T)$ have previously been observed in two other reports of the dielectric response of YMnO$_3$ [21,22], however, at different temperatures and with different amplitudes, which speaks against an intrinsic origin. In addition, during our measurements at 300 K < $T$ < 600 K, involving several heating/cooling cycles, at the first heating run the peak occurred at significantly higher temperature than at the subsequent runs [inset of Fig. 1(a)]. A change of oxygen stoichiometry at the sample surface seems a reasonable explanation for this phenomenon while it is unlikely that the stoichiometry of the whole sample changes at these relatively moderate temperatures [22]. Schottky diodes forming when metallic electrodes are applied to semiconducting samples are also often found to lead to MW relaxations [18,31,32]. To check for this possibility, we have performed measurements with different contact types [18]. They revealed only minor deviations [cf. squares and solid line in Fig. 1(a)] indicating that Schottky-diode formation only plays a minor role in our sample. Adem $et\ al.$ [23] recently reported a stronger influence of contact material on the giant relaxation mode found by them in YMnO$_3$, again pointing to the non-intrinsic and irreproducible nature of this surface-related spectral feature. In contrast to relaxation 2, we found the spectral features linked to relaxation 1 to be well reproducible in subsequent cooling and heating runs. We ascribe it to a MW relaxation caused by internal barrier layer capacitors (IBLCs) [33] arising from those parts of the ferroelectric DWs with low conductivity (the conducting parts of the DWs, also detected in [4], should not contribute to the dielectric properties).

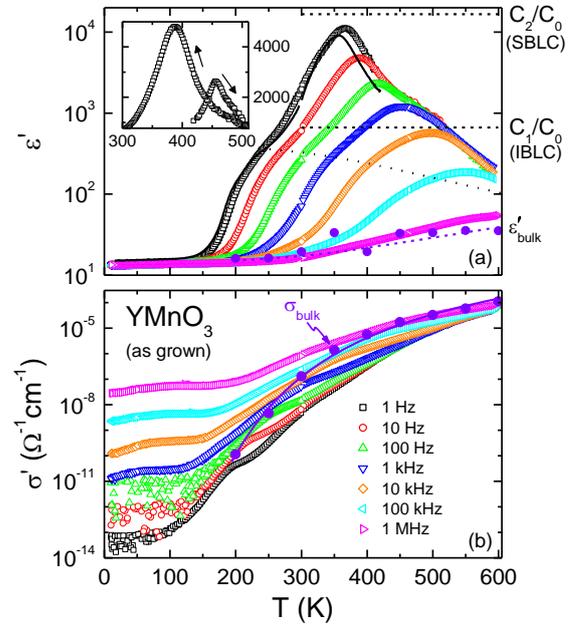

FIG. 1 (color online). Temperature dependence of (a) $\varepsilon'$ and (b) $\sigma'$ of sample 1 with silver-paint contacts measured for various frequencies under cooling (open symbols). The solid line in (a) shows $\varepsilon'(T)$ at 1 Hz for the same sample, using sputtered platinum contacts. The dotted line in (a) roughly indicates the separation of relaxations 1 and 2 showing up at low and high temperatures, respectively. The two horizontal dashed lines in (a) indicate the capacitances (divided by the empty capacitance $C_0$) of the two insulating regions used for the fits of the frequency-dependent data (Fig. 2). The closed circles in (a) represent $\varepsilon'_{bulk}$ as resulting from these fits; the corresponding dashed line is a guide for the eyes. The closed circles in (b) show the bulk dc conductivity as determined from the fits. The line in (b) is a fit of $\sigma_{bulk}$ with the Arrhenius law [cf. Fig. 3(b)]. The inset in (a) shows $\varepsilon'(T)$ at 1 Hz, measured at the first heating run and the subsequent cooling run.

Generally, thin layers within a sample, having lower conductivity than the bulk, act like capacitors (SBLCs or IBLCs) and lead to a MW relaxation and the typical signatures of relaxation processes, a step in $\varepsilon'(T,\nu)$ and a peak in $\varepsilon''(T,\nu)$. Possible causes are Schottky diodes or non-stoichiometric surface layers (SBLCs) and grain boundaries or other IBLCs. The static limit of the dielectric constant, $\varepsilon_{s,\mathrm{MW}}$, observed at low frequencies [for $\varepsilon'(\nu)$ plots] or high temperatures [for $\varepsilon'(T)$], is artificially enhanced compared to the bulk dielectric constant $\varepsilon'_{\mathrm{bulk}}$ by a factor given by the ratio of the sample thickness $d$ and overall layer thickness $D_l$, i.e., $\varepsilon_{s,\mathrm{MW}} \propto d/D_l$ [18] (note that $D_l$ is the summed-up, effective DW thickness and not the real thickness of a single DW). This is strictly valid if making the reasonable assumption that the intrinsic $\varepsilon'$ of the charge-carrier-depleted layers is of the same order as for the bulk. As $d/D_l$ can become very large, this explains the apparently colossal values of $\varepsilon'$ arising from surface effects in some materials [18,31,32,33] and in the present case of relaxation 2.

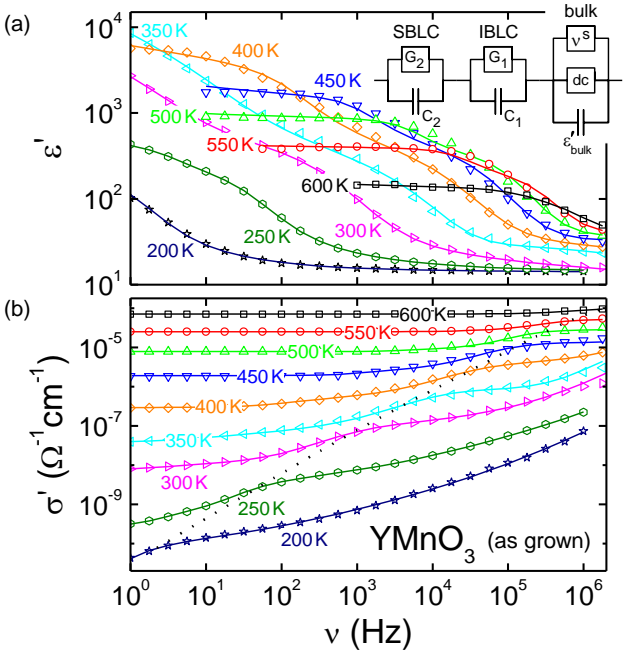

FIG. 2 (color online). (a) $\varepsilon'$ and (b) $\sigma'$ spectra of sample 1. The solid lines in (a) and (b) are fits using the equivalent circuit indicated in frame (a) (see text), simultaneously performed for $\varepsilon'(\nu)$ and $\sigma'(\nu)$. The dotted line in (b) connects the shoulders arising from relaxation 1.

To obtain quantitative information, we have analyzed the frequency dependence of the measured dielectric properties as shown for sample 1 in Fig. 2. Here in $\varepsilon'(\nu)$ [Fig. 2(a)] two successive steps are observed corresponding to relaxations 2 (at low frequencies) and 1 (high frequencies). The shoulders revealed in $\sigma'(\nu)$ [connected by the dotted line in Fig. 2(b)] arise from relaxation 1. At low temperatures and high frequencies, the bulk properties dominate the data (e.g., for 300 K above about 3 kHz) as the insulating thin layers, acting like lossy capacitors, become shortened [18,31]. For the bulk dielectric constant, $\varepsilon'_{\mathrm{bulk}} \approx 20$ is found. The bulk conductivity shows a nearly frequency-independent region signifying dc conductivity $\sigma_{\mathrm{dc}}$, followed by a power-law increase indicating hopping conductivity [18,34].

The lines in Fig. 2 are fits using the equivalent circuit shown in Fig. 2(a). The SBLCs and IBLCs, schematically indicated in Fig. 3(a), are both modeled by parallel RC circuits ($C_2 \| G_2$ and $C_1 \| G_1$, respectively, with $G$ the conductances) [18,31,33]. For the bulk part of the sample, a capacitor accounting for the intrinsic $\varepsilon'_{\mathrm{bulk}}$, a resistor for $\sigma_{\mathrm{dc}}$, and an element with a frequency-dependent conductivity $\sigma' \propto \sigma'' \propto \nu^s$ with $s < 1$ is used [18,31]. The latter corresponds to Jonscher's universal dielectric response [35] covering the mentioned increase of $\sigma'(\nu)$ due to hopping transport. Notably, perfect fits of the complete data set are possible with temperature-independent capacitors $C_1$ and $C_2$ [horizontal dashed lines in Fig. 1(a)] as usually found for SBLCs and IBLCs [18,31,32,33]. $\varepsilon'_{\mathrm{bulk}}(T)$ as resulting from the fits is included in Fig. 1(a) (closed circles). It varies between about 15 and 35. Obviously, this equivalent-circuit approach provides a good description of our data without assuming contributions from segmental DW oscillations found in some canonical ferroelectrics [36].

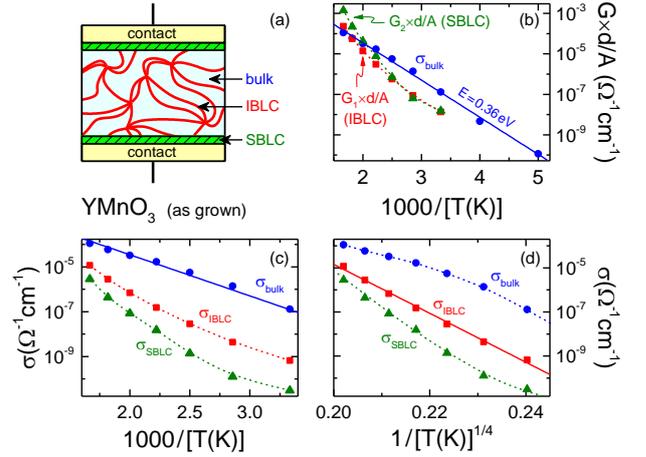

FIG. 3 (color online). (a) Schematic sample cross section (not to scale) indicating bulk, SBLC, and IBLC regions, where the IBLCs arise from the DWs (this picture is somewhat oversimplified as only part of the DWs have low conductivity [4]). (b) Arrhenius representation of the bulk dc conductivity and apparent layer conductivities (conductances multiplied by the bulk geometry-factor $d/A$) of sample 1 as obtained from the fits shown in Fig. 2. (c) Bulk dc conductivity and true layer conductivities calculated as described in the text. (d) Same as (c) but using a representation that linearizes $\sigma(T)$ for variable range hopping. The solid lines in (b), (c), and (d) are linear fits; the dashed lines are guides to the eye.

In the following, we deduce the intrinsic conductivities of the DWs and SBLCs from the fit results. Figure 3(b) shows the apparent conductivities obtained from the fits, which correspond to the conductances of the equivalent circuit multiplied by the bulk geometry factor $d/A$ ($A$: sample area). However, only for the bulk this represents the *intrinsic* con-ductivity $\sigma_{bulk}$. The latter is also included in Fig. 1(b) revealing the typical behavior for MW relaxations [18,37,38]. $\sigma_{bulk}$ follows the Arrhenius law with an activation energy of 0.36 eV. The same value was also found at the highest frequencies and lowest temperatures investigated in Ref. [21], where obviously the intrinsic bulk response was detected. Figure 3(b) reveals that the layer conductances cross the bulk curve at high temperatures. Such behavior can cause an apparent relaxor-ferroelectric signature in $\varepsilon'(T,\nu)$ as indeed observed in Fig. 1 [18,29].

In contrast to the bulk, for the layers the quantity $G \times d/A$ plotted in Fig. 3(b) only represents the *apparent* conductivity because their overall thickness $D_l$ is smaller than the sample thickness $d$. To calculate the *intrinsic* layer conductivities $\sigma_{SBLC}$ and $\sigma_{IBLC}$, $G \times d/A$ shown in Fig. 3(b) has to be divided by $d/D_l$. This ratio can be estimated from our fit results if considering the mentioned enhancement of $\varepsilon_{s,MW}$ of the corresponding MW relaxation, which is determined by the ratio of sample and overall layer thickness, i.e., $\varepsilon'_{s,MW}/\varepsilon'_{bulk} = d/D_l$. This relation is valid if making the reasonable assumption of homogeneity of $\varepsilon'$ all over the sample. Using an average value of $\varepsilon'_{bulk} \approx 20$ and the $\varepsilon_{s,MW}$ values obtained from the fits, this ratio is about 33 for the IBLCs and 830 for the SBLCs. The intrinsic layer conductivities calculated using these values are shown in Fig. 3(c), together with $\sigma_{bulk}$. $\sigma_{SBLC}$ is by about 2 - 4 decades lower than $\sigma_{bulk}$. Of special interest is the result for $\sigma_{IBLC}$ revealing that, at room temperature, the conductivity of the insulating DWs is by more than two decades lower than for the bulk ($\sigma_{bulk}/\sigma_{IBLC} \approx 190$). Thus the conductivity contrast is significantly stronger than the conductance ratios obtained for two related hexagonal manganites from c-AFM measurements [4,8], which do not reveal absolute values of the conductivities. It should be noted that the obtained DW conductivity is a representative value of the insulating parts of the DWs and our results do not exclude conductivity variations in dependence of polarization orientation as detected in [4] and [8].

Figure 3(c) demonstrates clear deviations of $\sigma_{IBLC}(T)$ from thermally activated behavior. Instead it follows $\sigma_{IBLC} \sim \exp[-(T_0/T)^{1/4}]$, as predicted for variable range hopping [Fig. 3(d)] pointing to phonon-assisted tunneling of Anderson-localized electrons or holes [39]. In Ref. [16], for ErMnO$_3$ it was explicitly demonstrated that the insulating DWs are depleted from mobile charge carriers (holes). From our results we conclude that the charge transport in these DWs is dominated by tunneling of the remaining localized charge carriers. Interestingly, recent electron-energy loss spectroscopy and density-functional theory calculations suggest charge transport via minority charge carriers (electrons) for the insulating DWs at sufficiently large voltage [40]. Their localized nature and small carrier density of about 0.1 per Mn ion is well consistent with the occurrence of hopping charge transport as found in the present work.

As shown before [24,25], annealing hexagonal manganites above $T_c$ and subsequent cooling with different rates leads to well-defined vortex densities which depend on cooling rate.

The right part of Fig. 4 presents PFM images of the annealed samples 1a and 2, exhibiting the typical domain patterns [2,4,24]. A vortex density $\rho_v$ of about $4\times10^4$/mm$^2$ is deduced for sample 1a, while sample 2 has significantly higher $\rho_v$ of about $1.5\times10^6$/mm$^2$. The dielectric properties of these samples also reveal the presence of two relaxation processes, which were interpreted and analyzed in the same way as for sample 1 [29]. The obtained intrinsic conductivities are shown in Fig. 4. For the annealed sample 1a, the bulk conductivity at room temperature is by more than two decades higher than for the as-grown state of this sample and a lower energy barrier of 0.26 eV is found. Most interestingly, we obtain a conductivity contrast between bulk and insulating DWs of about 500 at room temperature, even higher than in the untempered sample. In marked contrast, while sample 2 exhibits a bulk conductivity and energy barrier of similar order as for the untempered sample 1, it seems to have a much smaller conductivity contrast of about 1.3 only at room temperature. Different annealing/cooling cycles obviously have a pronounced effect on the MW relaxations in YMnO$_3$ [29] pointing to marked variations of bulk and DW conductivity. Moreover, the different crystal-growth procedure of sample 2 may also play a role for its different behavior. Further work is necessary to systematically investigate the dependence of the intrinsic conductivities in the hexagonal manganites on sample history.

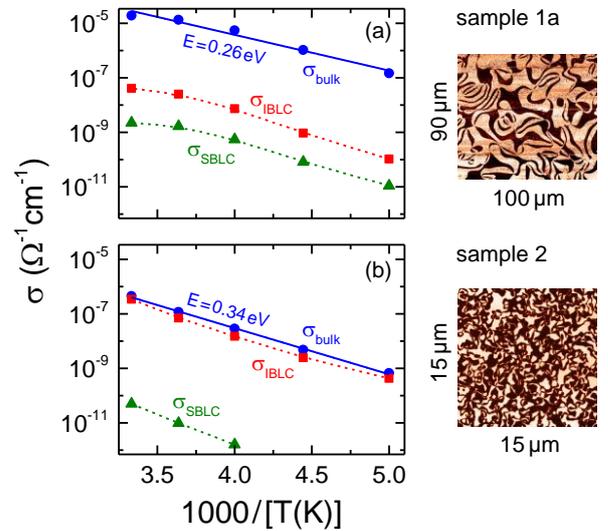

FIG. 4 (color online). Bulk dc conductivity and true layer conductivities of the annealed samples 1a (a) and 2 (b), deduced from the equivalent-circuit analysis as described in the text. The solid lines in both frames are linear fits; the dashed lines are guides to the eye. At the right side of the figures, PFM images of the *ab* sample surfaces are shown.

In summary, we have measured the dielectric properties of single-crystalline YMnO$_3$ and performed a detailed equivalent-circuit analysis of the obtained spectra. From this, we were able to determine absolute values of the conductivity of the ferroelectric DWs with suppressed conductance and of the conductivity contrast between the domains and DWs. For the latter, we find values up to 500 at room temperature and indications for a strong dependence on the thermal history and crystal-growth procedure of the samples. The charge transport in the insulating DWs is dominated by the tunneling of localized charge carriers. These results demonstrate that dielectric spectroscopy can disclose the intrinsic conductivity properties of ferroelectric DWs, which is prerequisite for the development of new domain-wall-based nanotechnology.

We thank N. A. Spaldin for stimulating discussions. This work was supported by the Deutsche Forschungsgemeinschaft through the Transregional Collaborative Research Center TRR 80 (Augsburg, Munich, Stuttgart), by the BMBF via ENREKON 03EK3015, and by the SNSF Grant No. 200021_149192/1.

---

*Corresponding author. Peter.Lunkenheimer@Physik.Uni-Augsburg.de

[1] B. B. Van Aken, T. T. M. Palstra, A. Filippetti, and N. A. Spaldin, Nat. Mater. **3**, 164 (2004).
[2] T. Choi, Y. Horibe, H. T. Yi, Y. J. Choi, W. Wu, and S.-W. Cheong, Nat. Mater. **9**, 253 (2010).
[3] M. Mostovoy, Nat. Mater. **9**, 188 (2010).
[4] D. Meier, J. Seidel, A. Cano, K. Delaney, Y. Kumagai, M. Mostovoy, N. A. Spaldin, R. Ramesh, and M. Fiebig, Nat. Mater. **11**, 284 (2012).
[5] M. Lilienblum, T. Lottermoser, S. Manz, S. M. Selbach, A. Cano, and M. Fiebig, Nat. Phys. **11**, 1070 (2015).
[6] T. Jungk, Á. Hoffmann, M. Fiebig, and E. Soergel, Appl. Phys. Lett. **97**, 012904 (2010).
[7] M. Fiebig, Th. Lottermoser, D. Fröhlich, A. V. Goltsev, and R. V. Pisarev, Nature (London) **419**, 818 (2002).
[8] W. Wu, Y. Horibe, N. Lee, S.-W. Cheong, and J. R. Guest, Phys. Rev. Lett. **108**, 077203 (2012).
[9] E. Salje and H. Zhang, Phase Transitions **82**, 452 (2009).
[10] H. Béa and P. Paruch, Nat. Mater. **8**, 168 (2009).
[11] G. Catalan, J. Seidel, R. Ramesh, and J. F. Scott, Rev. Mod. Phys. **84**, 119 (2012).
[12] M. Fiebig, Phil. Trans. R. Soc. A **370**, 4972 (2012).
[13] D. Meier, J. Phys.: Condens. Matter **27**, 463003 (2015).
[14] C. Becher, L. Maurel, U. Aschauer, M. Lilienblum, C. Magén, D. Meier, E. Langenberg, M. Trassin, J. Blasco, I. P. Krug, P. A. Algarabel, N. A. Spaldin, J. A. Pardo, and M. Fiebig, Nat. Nanotechnol. **10**, 661 (2015).
[15] Y. S. Oh, X. Luo, F.-T. Huang, Y. Wang, and S.-W. Cheong, Nat. Mater. **14**, 407 (2015).
[16] J. Schaab, A. Cano, M. Lilienblum, Z. Yan, E. Bourret, R. Ramesh, M. Fiebig, and D. Meier, Adv. Electron. Mater. **2**, 1500195 (2016).
[17] J. Schaab, I. P. Krug, F. Nickel, D. M. Gottlob, H. Doğanay, A. Cano, M. Hentschel, Z. Yan, E. Bourret, C. M. Schneider, R. Ramesh, and D. Meier, Appl. Phys. Lett. **104**, 232904 (2014).
[18] P. Lunkenheimer, S. Krohns, S. Riegg, S. G. Ebbinghaus, A. Reller, and A. Loidl, Europhys. J. Spec. Top. **180**, 61 (2010).
[19] V. Bobnar, P. Lunkenheimer, M. Paraskevopoulos, and A. Loidl, Phys. Rev. B **65**, 184403 (2002).
[20] I. G. Ismailzade and S. A. Kizhaev, Sov. Phys. Solid State **7**, 236-238 (1965).
[21] M. Tomczyk, P. M. Vilarinho, A. Moreira, and A. Almeida, J. Appl. Phys. **110**, 064116 (2011).
[22] P. Ren, H. Fan, and X. Wang, Appl. Phys. Lett. **103**, 152905 (2013).
[23] U. Adem, N. Mufti, A. A. Nugroho, G. Catalan, B. Noheda, and T. T. M. Palstra, J. Alloys Compd. **638**, 228 (2015).
[24] S. C. Chae, N. Lee, Y. Horibe, M. Tanimura, S. Mori, B. Gao, S. Carr, and S.-W. Cheong, Phys. Rev. Lett. **108**, 167603 (2012).
[25] S. M. Griffin, M. Lilienblum, K. T. Delaney, Y. Kumagai, M. Fiebig, and N. A. Spaldin, Phys. Rev. X **2**, 041022 (2012).
[26] L. E. Cross, Ferroelectrics **76**, 241 (1987).
[27] A. Levstik, Z. Kutnjak, C. Filipič, and R. Pirc, Phys. Rev. B **57**, 11204 (1998).
[28] G. Catalan, D. O'Neill, R. M. Bowman, and J. M. Gregg, Appl. Phys. Lett. **77**, 3078 (2000).
[29] See Supplemental Material at [URL will be inserted by publisher] for more details on the non-intrinsic relaxor behavior and the dielectric results for samples 1a and 2.
[30] S. Krohns, P. Lunkenheimer, S. G. Ebbinghaus, and A. Loidl, Appl. Phys. Lett. **91**, 022910 (2007).
[31] P. Lunkenheimer, V. Bobnar, A. V. Pronin, A.I. Ritus, A. A. Volkov, and A. Loidl, Phys. Rev. B **66**, 052105 (2002).
[32] P. Lunkenheimer, R. Fichtl, S. G. Ebbinghaus, and A. Loidl, Phys. Rev. B **70**, 172102 (2004).
[33] D. C. Sinclair, T. B. Adams, F. D. Morrison, and A. R. West, Appl. Phys. Lett. **80**, 2153 (2002).
[34] S. R. Elliott, Adv. Phys. **36**, 135 (1987).
[35] A. K. Jonscher, Nature (London) **267**, 673 (1977).
[36] W. Kleemann, Annu. Rev. Mater. Res. **37**, 415 (2007)
[37] A. Seeger, P. Lunkenheimer, J. Hemberger, A. A. Mukhin, V. Yu. Ivanov, A. M. Balbashov, and A. Loidl, J. Phys.: Condens. Matter **11**, 3273 (1999).
[38] A. Ruff, S. Krohns, F. Schrettle, V. Tsurkan, P. Lunkenheimer, and A. Loidl, Eur. Phys. J. B **85**, 290 (2012).
[39] N. F. Mott and E. A. Davis, *Electronic Processes in Non-Crystalline Materials* (Clarendon Press, Oxford, 1979).
[40] J. A. Mundy, J. Schaab, Y. Kumagai, A. Cano, M. Stengel, I. P. Krug, D. M. Gottlob, H. Doganay, M. E. Holtz, R. Held, Z. Yan, E. Bourret, C. M. Schneider, D. G. Schlom, D. A. Muller, R. Ramesh, N. A. Spaldin, and D. Meier, unpublished.

# Supplemental Material

# Conductivity contrast and tunneling charge transport in the vortex-like ferroelectric domain patterns of multiferroic hexagonal YMnO$_3$

E. Ruff[1], S. Krohns[1], M. Lilienblum[2], D. Meier[2,3], M. Fiebig[2], P. Lunkenheimer[1,*], A. Loidl[1]

[1]Experimental Physics V, Center for Electronic Correlations and Magnetism, University of Augsburg, 86159 Augsburg, Germany
[2]Department of Materials, ETH Zürich, 8093 Zürich, Switzerland
[3]Department of Materials Science and Engineering, Norwegian University of Science and Technology, 7043 Trondheim, Norway

*e-mail: Peter.Lunkenheimer@Physik.Uni-Augsburg.de

## 1. Non-intrinsic relaxor behavior

Here we explain in more detail how the relaxor behavior observed for relaxation 2 in the as-grown sample 1 of YMnO$_3$ can arise from non-intrinsic effects. These considerations are based on Refs. [16] and [26] of the main paper. As mentioned in the main text, Maxwell-Wagner (MW) relaxations can be caused by the existence of thin insulating layers within the sample or at its surface, termed internal barrier layer capacitors (IBLCs) or surface barrier layer capacitors (SBLCs), respectively. Possible origins were discussed in the main text (see also Ref. 16). As mentioned there, then the sample can be described by a parallel RC circuit for the layers (with $R_l$ and $C_l$ for the layer resistance and capacitance, respectively), connected in series to the bulk sample [Fig. S1(a)].

For simplicity reasons, here we assume that there is only one type of insulating layers instead of the two layers arising from IBLCs and SBLCs in the actual YMnO$_3$ samples. As discussed in the main text, we ascribe the relaxor-like MW relaxation to the SBLCs and, thus, the quantities $G_2$ and $C_2$ assigned to the SBLCs in the main paper are related to $R_l$ and $C_l$ via $R_l = 1/G_2$ and $C_l = C_2$. Moreover, we neglect here the $\sigma' \propto \nu^s$ contribution from hopping conductivity to the bulk response, i.e., the sample is modeled by a simple parallel circuit of $R_b$ and $C_b$ [Fig. S1(a)]. Without these assumptions, the formulae describing the equivalent circuit become exceedingly complex while the general conclusions are the same.

The total admittance (complex conductance) of a single RC circuit is $Y = G + i\omega C$, with the conductance $G = 1/R$ and $\omega = 2\pi\nu$. The circuit shown in Fig. S1(a) has a total admittance of [16]

$$Y_{tot} = G'_{tot} + iG''_{tot} = \frac{Y_l Y_b}{Y_l + Y_b} = \frac{(G_l + i\omega C_l)(G_b + i\omega C_b)}{G_l + G_b + i\omega(C_l + C_b)}.$$

(Here the indices "$l$" and "$b$" stand for "layer" and "bulk", respectively.) Resolving this into real and imaginary part and calculating the capacitance via $C'_{tot} = G''_{tot}/\omega$ leads to [16]:

$$C'_{tot} = \frac{\left(G_l^2 C_b + G_b^2 C_l\right) + \omega^2 C_l C_b (C_l + C_b)}{(G_l + G_b)^2 + \omega^2 (C_l + C_b)^2} \quad (S1)$$

As noted, e.g., in Ref. [16], this formally results in exactly the same frequency dependence as for a Debye relaxation with some additional loss contribution due to dc conductivity. Such non-intrinsic relaxational response is commonly termed Maxwell-Wagner relaxation.

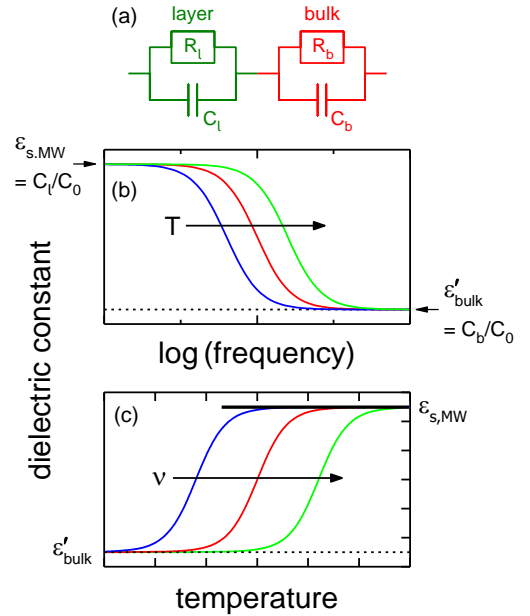

FIG. S1. (a) Simplified equivalent circuit for a heterogeneous sample consisting of a bulk and a layer contribution. (b) Schematic plot of the frequency dependence of the dielectric constant for a conventional MW relaxation. (c) Schematic plot of the temperature dependence. The horizontal solid line indicates a temperature independent $\varepsilon_{s,\mathrm{MW}}$ as expected for this case.

Here we are interested in the capacitance for $\nu \to 0$, $C_{s,\mathrm{MW}}$, which in YMnO$_3$ shows an increase with decreasing temperature, typical for relaxor ferroelectrics [solid line in Fig. S2(a)]. From Eq. (S1), one obtains [16]:

$$C_{s,\mathrm{MW}} = \frac{G_l^2 C_b + G_b^2 C_l}{(G_l + G_b)^2} \quad (S2)$$

From $C_{s,\text{MW}}$ the static dielectric constant $\varepsilon_{s,\text{MW}}$ of the MW relaxation can be calculated via $\varepsilon_{s,\text{MW}} = C_{s,\text{MW}}/C_0$ where $C_0$ is the empty capacitance deduced from the sample geometry.

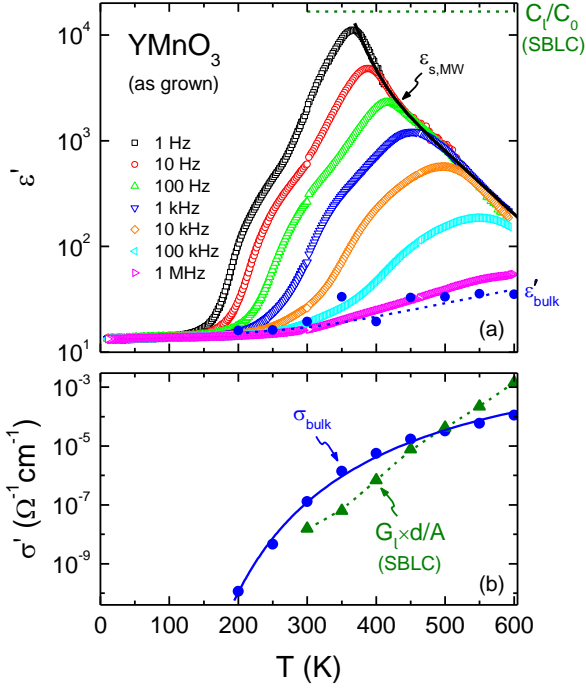

FIG. S2. (a) Temperature dependence of $\varepsilon'$ as measured at different frequencies [same data as shown in Fig. 1(a)]. The closed circles show the intrinsic bulk dielectric constant as deduced from the fits shown in Fig. 2; the corresponding dashed line is a guide for the eyes. The solid line indicates $\varepsilon_{s,\text{MW}}(T)$. (b) Temperature dependence of the bulk conductivity and the apparent layer conductivity arising from the SBLCs (conductance multiplied by the bulk geometry factor $d/A$) of sample 1 as obtained from the fits shown in Fig. 2 [cf. Fig. 3(b), where the same data are shown in Arrhenius representation]. The solid line represents an Arrhenius fit (see main paper). The dashed line is a guide for the eyes.

In the simplest and most common case, the overall conductance of the insulating layers is much lower and their capacitance is much higher than that of the bulk sample, i.e., $G_l \ll G_b$ and $C_l \gg C_b$. According to Eq. (S2), the static capacitance is then given by $C_{s,\text{MW}} = C_l$ and $\varepsilon_{s,\text{MW}} = C_l/C_0$. For this scenario, obviously the layer properties completely dominate the detected dielectric response at low frequencies or high temperatures [Figs. S1(b) or (c), respectively] [16]. In contrast, at high frequencies or low temperatures, the response is dominated by the bulk properties because, with increasing frequency (or decreasing temperature), the layer RC-circuit becomes successively shorted [16].

The layer capacitance $C_l$ (and thus $C_{s,\text{MW}}$) usually is temperature independent. For example, in depletion layers of Schottky diodes the dielectric constant is determined by the nearly temperature-independent ionic and electronic polarizability. The same is valid for a non-stoichiometric surface layer as considered here for the large relaxation 2 observed in YMnO$_3$ (see main text). Therefore, for most MW relaxations the temperature dependence of $\varepsilon'$ follows the behavior indicated in Fig. S1(c), with a nearly temperature-independent $\varepsilon_{s,\text{MW}}$ (horizontal solid line) and no indication of relaxor-ferroelectric behavior (for examples, see Refs. [16,28,29]).

However, if one of the conditions $G_l \ll G_b$ and $C_l \gg C_b$ no longer is valid, the situation is more complex. This becomes immediately obvious from Eq. (S2): Bulk and layer conductivity usually have semiconducting temperature characteristics and thus $G_b$ and $G_l$ are strongly temperature dependent. If the conditions $G_l \ll G_b$ or $C_l \gg C_b$ are invalid, this can lead to considerable temperature dependence of $C_{s,\text{MW}}$ and $\varepsilon_{s,\text{MW}}$. As revealed by the upper dashed line in Fig. 1(a) of the main paper and in Fig. S2(a), $C_l \gg C_b$ is always fulfilled for the SBLC relaxation in YMnO$_3$: This dashed line indicates $C_l/C_0$ while the closed circles show $\varepsilon'_{\text{bulk}} = C_b/C_0$, i.e., $C_l$ is by about three decades larger than $C_b$. However, the condition $G_l \ll G_b$ is not fulfilled in the whole temperature range: Figure S2(b) compares the bulk dc conductivity $\sigma_{\text{bulk}} = G_b \times d/A$ ($d$: sample thickness, $A$: sample area) with the apparent SBLC conductivity $G_l \times d/A$ as obtained from the fits shown in Fig. 2. Obviously both curves cross. In the temperature region around 350 K, $G_b$ is by about a factor 20 larger than $G_l$. Thus, $G_l \ll G_b$ is approximately fulfilled and the detected high $\varepsilon'$ values of the order of $10^4$ are to a large extent dominated by the SBLCs. However, above about 350 K, both conductances approach each other and become equal at about 490 K. There $C_l \gg C_b$ and $G_l \approx G_b$ is valid and Eq. (S2) results in $C_{s,\text{MW}} = C_c/4$, predicting a considerable reduction with increasing temperature. Moreover, for $T > 490$ K, $G_l$ becomes even larger than $G_b$, finally exceeding it by about one decade at 600 K [Figure S2(b)]. For further increasing temperatures, in Eq. (S2) finally the condition $G_l \gg G_b$ will dominate over the relation $C_l \gg C_b$ (the conductances are squared in the denominator). Then $C_{s,\text{MW}} = C_b$ is expected, i.e., the static capacitance (and $\varepsilon_{s,\text{MW}} = C_{s,\text{MW}}/C_0$) becomes even more reduced with increasing temperature. This explains the observed relaxor-like decrease of $\varepsilon_{s,\text{MW}}$ with increasing temperature above about 350 K as revealed in Figs. 1(a) and S2(a).

One may note that, at 490 K, $\varepsilon_{s,\text{MW}}$ is actually reduced (compared to $C_l/C_0$; upper dashed line) by more than the predicted factor of four in Fig. S2(a). We ascribe this to the simplified equivalent circuit used here [Fig. S1(a)], which neglects the presence of the IBLC relaxation and the $v^s$ conductivity contribution for the bulk. Indeed, as demonstrated in the main paper, the experimental data can be well fitted with the complete circuit using a $C_l$ that is temperature independent over the whole temperature range.

In summary, the observed strongly temperature-dependent static dielectric constant of YMnO$_3$, apparently resembling relaxor-ferroelectric behavior, can be explained by an equivalent circuit leading to a MW relaxation. Within this scenario, the temperature dependence of $\varepsilon_{s,\text{MW}}$ arises from the crossing of the conductances $G_b(T)$ and $G_l(T)$ revealed in Fig. S2(b) and Fig. 3(b). It should be noted that the *conductivity* of the SBLCs of course is much lower than that of the bulk at all temperatures as revealed in Fig. 3(c) of the main paper. However, these layers obviously are relatively thin and therefore their *conductance* exceeds that of the bulk, at least at high temperatures [Fig. S2(b)].

## 2. Dielectric properties of samples 1a and 2

Figure S3 shows the temperature dependence of $\varepsilon'$ (a) and the frequency dependence of $\varepsilon'$ (b) and of $\sigma'$ (c) of sample 1a. Two relaxation processes are clearly revealed. The corresponding results for sample 2 are shown in Fig. S4. Here the smaller relaxation shows up as weak smeared-out step at the onset of the large relaxation step 2. The lines in Figs. S3(b) and (c) and S4(b) and (c) are fits with the equivalent circuit as described for sample 1 in the main text. The obtained intrinsic conductivities are shown in Fig. 4 of the main paper. Of course, one should be aware that the quantification of the DW conductivity as shown in Fig. 4 is based on an assignment of the MW relaxations in samples 1a and 2, analogous to that used for sample 1.

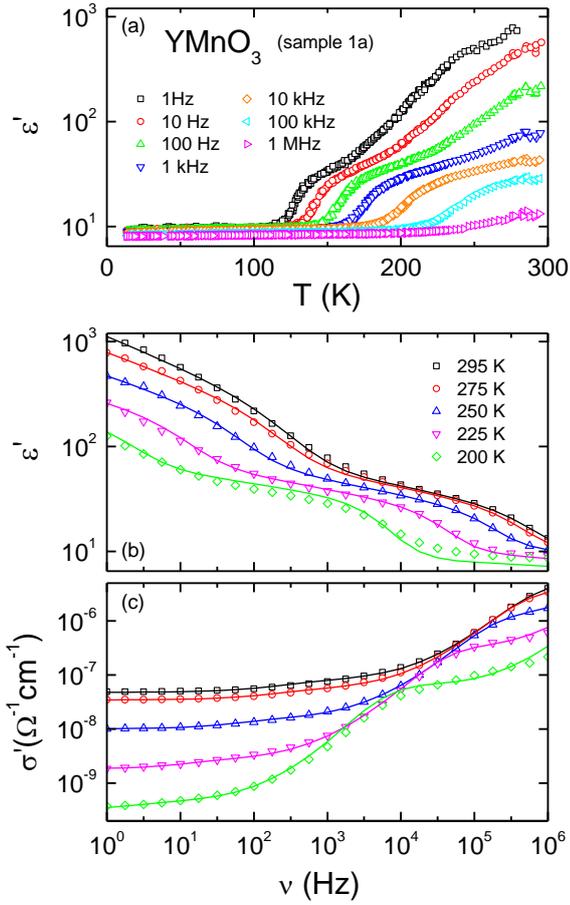

FIG. S3. Temperature dependence of $\varepsilon'$ (a) and frequency dependence of $\varepsilon'$ (b) and of $\sigma'$ (c) of sample 1a with silver-paint contacts. The solid lines in (b) and (c) show fits using the equivalent circuit indicated in the inset of Fig. 2(a).

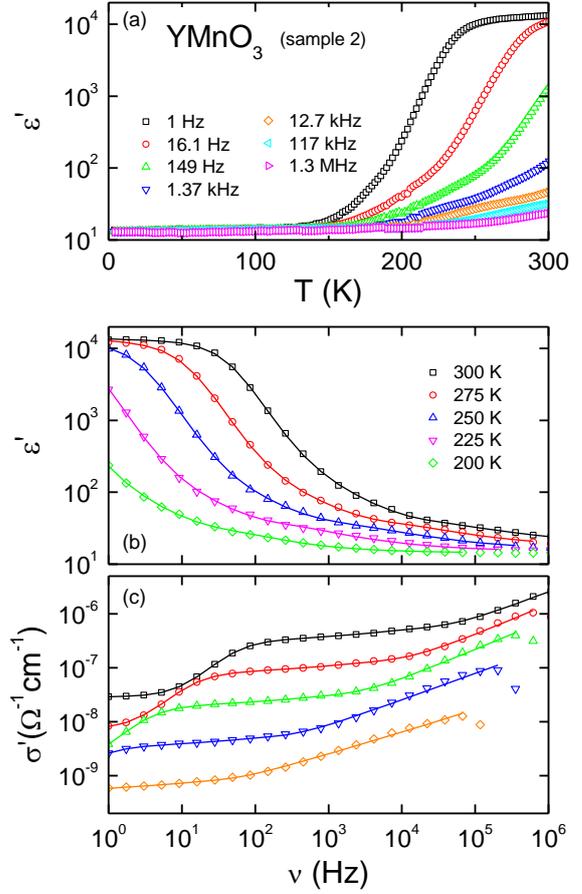

FIG. S4. Temperature dependence of $\varepsilon'$ (a) and frequency dependence of $\varepsilon'$ (b) and of $\sigma'$ (c) of sample 2 with silver-paint contacts. The solid lines in (b) and (c) show fits using the equivalent circuit indicated in the inset of Fig. 2(a).